\documentclass[twocolumn,secnumarabic,amssymb,amsmath, nobibnotes, aps, prl]{revtex4-1}

\usepackage{bm}

\usepackage{graphicx}
\usepackage{amsmath}

\begin{document}

\title{Giant Compton Shifts in Hyperbolic Metamaterials}

\author{Ivan V.~Iorsh$^{1}$}


\author{Alexander N. Poddubny$^{2}$}

\author{Pavel Ginzburg$^{1}$}

\author{Pavel A.~Belov$^{1}$}%

\author{Yuri S. Kivshar$^{1,3}$}

\affiliation {$^1$ITMO University, St. Petersburg 197101, Russia}
\affiliation{$^2$Ioffe Physical-Technical Institute, Russian Academy of Science, St.~Petersburg 194021, Russia}
\affiliation {$^3$Nonlinear Physics Center and CUDOS@ANU, Australian National University, Canberra ACT 0200, Australia}

\begin{abstract}
We study the Compton scattering of light by free electrons inside a hyperbolic medium. We demonstrate that the unconventional dispersion and local density of states of the electromagnetic modes in such media can lead to a giant Compton shift and dramatic enhancement of the scattering cross section. We develop an universal approach for the study of coupled multi-photon processes in nanostructured media and derive the spectral intensity function of the scattered radiation for realistic metamaterial structures.  We predict the Compton shift of the order of a few meVs for the infrared spectrum that is at least one order of magnitude larger than the Compton shift in any other system. 
\end{abstract}

\maketitle

Scattering is the most fundamental process enabling to probe internal structure of matter~\cite{book}. The principle of scattering experiments is to monitor a deviation of particles (fermions or bosons) from their original unperturbed trajectories and then collect the information about a scattering potential. One of the pioneering scattering experiments was performed by Lord Rutherford for the discovery of a structure of atoms~\cite{book2} extracted from the $\alpha$-particle scattering from a gold target. More recently, the scattering experiments became the most frequently used tools in atomic physics and related cross-disciplinary areas. Indeed, tailoring and manipulating various scattering processes play a fundamental role in modern physics.

One of the main parameters defining probability of any scattering process is the density of available states, which a scattered particle could occupy. While under most frequently used circumstances this parameter is predefined by a scattering object, the surrounding environment could contribute significantly to the process. This idea is most frequently employed for manipulation of spontaneous light radiation by introducing an emitter into a cavity~\cite{Purcell}.

\begin{figure}[!h]
\centerline{\includegraphics[width = 0.8\columnwidth]{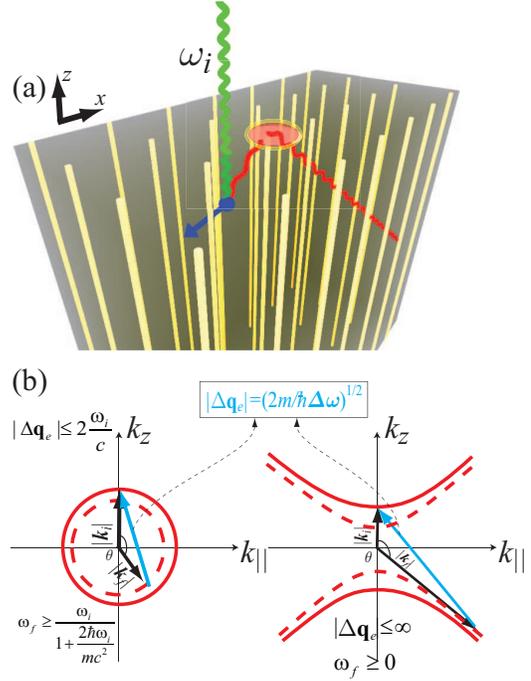}}
\caption{(Color online) (a) Schematic of the Compton scattering in a hyperbolic medium. An incident photon (green)
is scattered (red) by a free electron (blue). (b) Diagrams illustrating the scattering process in vacuum (left) and in a hyperbolic medium (right). Red curves correspond to the initial (solid) and scattered (dashed) photon isofrequency surfaces, blue vector shows the momentum of the electron after scattering. }
\label{fig_1}
\end{figure}

However, this concept could be pushed even further and broadband non-resonant engineering of local density of states (LDOS) could be performed with the help of artificially created structures, namely metamaterials. In this context, hyperbolic metamaterials will play a key role. Hyperbolic metamaterials are  a special class of  media~\cite{smith2003} with electromagnetic properties described by the diagonal permittivity tensor with the principal components being of the opposite signs which results in a hyperbolic shape of the isofrequency contours~\cite{narimanov2010,Rev1}.  Realizations of such substances could be based on metal-dielectric layered structures~\cite{layeredhmm}, arrays of vertically aligned metal nanorods~\cite{nanorod}, or semiconductor heterostructures~\cite{semicond,Ginz1}.  One of the most intriguing properties of the hyperbolic media is the non-resonant broadband enhancement of LDOS, (limited only by a smallest yet finite system's dimension~\cite{micros}), affecting emitters, situated inside or in the vicinity of the hyperbolic metamaterial.

While spontaneous emission and photon absorption are the first order perturbation process and directly influenced by LDOS, higher order interactions could have nontrivial dependence on it and, as the result, be strongly tailored by electromagnetic environment~\cite{NL2hyp, NL1hyp,2ph}. In this Letter we investigate the influence of LDOS on higher order light-matter interaction processes and, in particular, concentrate on Compton scattering~\cite{XRAYrev}, which was shown to be significantly modified by hyperbolic media. First we show that commonly used theoretical quantum methods~\cite{Greiner} result in unphysical divergences of both frequency shifts and scattering cross-sections and, as the result, are inapplicable in the case of metamaterial assembly. Relying on the demand to resolve this type of scattering catastrophe, a  quantum formalism, based on Langevin approach, was developed. Proper account for the inherent losses and finite size of the metamaterial sample led to finite, yet giant frequency shifts and cross-section enhancement, making hyperbolic metamaterials to hold a promise to enable new type of nonlinear interactions.

In the classical Compton scattering setup the incident photon is spontaneously re-emitted in arbitrary direction (with variable probability), while the free electron compensates over the momentum and energy mismatches between incident and scattered quanta~\cite{Greiner}. In the case of finite nanostructures and, in particular when polaritonic waves propagating inside metamaterials are involved, this description fails to represent the physical phenomenon and should be reconsidered with an emphasis on photonic structure details.

We consider the set-up shown in Fig.~\ref{fig_1}(a), where a photon of a given frequency is normally incident upon the semi-infinite hyperbolic media. The presence of the interface enables to approach a probable experimental layout as well as it highlights the impact of geometric arrangements. It is worth noting, that arbitrary shaped object could be treated within the same approach, nevertheless numerical evaluation of dispersion relations will be required.

The incident free space photon was chosen to propagate along the extraordinary direction (along which the permittivity is negative) and polarized along the positive $\varepsilon$, as it will allow the wave to propagate inside the hyperbolic bulk and not being reflected.   Within our theoretical approach, which utilizes the Born approximation,  the electrons in the conducting wires can not play the role of the scattering centers since their interaction with electromagnetic field is accounted for in the effective negative dielectric permittivity of the wires. We thus need to generate an additional population of the free electrons in the system, which can be achieved using pump-probe approach. Namely, we should first excite free electron population with a UV pump pulse due to the photoelectric effect and then consider the scattering of the probe IR pulse by these free electrons.

The essence of this effect can be illustrated using the scattering diagram shown in Fig.~\ref{fig_1}(b). Momentum conservation implies that the acquired electron momentum $\Delta \mathbf{q}_e$ is equal to the difference of initial and final photon wavevectors.  {In vacuum the wave vector of photon is much smaller than the wave vector of electron of the same energy. Hence, as can be seen in Fig.~\ref{fig_1}(b), the maximum electron wave vector $|\Delta \mathbf{q}_e|$ is limited by $2\omega_i/c$. This} defines the upper bound for the electron kinetic energy, that is equal to the photon energy shift due to the energy conservation. In {sharp} contrast to the vacuum case, in hyperbolic medium the final photon momentum can be arbitrarily large, leading to the possibility for the large electron kinetic energy and thus the large Compton frequency shift. The scattered quanta then occupies {one} of  the large wave vector states, TM  eigenmode. It is wroth noting, that TM eigenmodes, having $k$-vector much longer than those in the free space, will experience total internal reflection and, as the result, will be trapped inside the hyperbolic bulk. However, its near field (schematically shown with the glowing disc on Fig. 1(a)) can be detected within the near field close to the interface or directly inside the hyperbolic metamaterial. Alternatively, the surface of the metamaterial could be patterned to enhance the coupling between the eigenmodes of hyperbolic medium and propagating waves in vacuum~\cite{nanopatt}. A proper theoretical description of the above scenario will be developed hereafter. The permittivity tensor of the metamaterial block is given by: $\hat{\varepsilon}=\mathrm{diag}[\varepsilon_{xx},\varepsilon_{xx},\varepsilon_{zz}]$.
Transverse magnetic (TM) modes  have high LDOS and hence will be considered hereafter. The dispersion relation for those modes is given by: $(\omega/c)^2=k_x^2/\varepsilon_{zz}+k_z^2/\varepsilon_{xx}$.

In order to calculate the frequency shift and the differential cross-section of the scattering process, we introduce the Hamiltonian for the electron system coupled to the electromagnetic field~\cite{Welsch,Bruus}:
\begin{align}
&\hat{H}=\hat{H}^{el}_{\mathrm{kin}}+\hat{H}^{phot}_{\mathrm{kin}}+\frac{e}{c}\int d\mathbf{r}\hat{\mathbf{J}}\hat{\mathbf{A}},\label{hnoloss}\\
&\hat{\mathbf{J}}=\frac{2\hbar}{m}\displaystyle\sum_{\mathbf{k}\mathbf{q}}(\mathbf{k}+\frac{\mathbf{q}}{2})e^{i\mathbf{qr}}\hat{c}^{\dagger}_{\mathbf{k}}\hat{c}_{\mathbf{k+q}}+\frac{e}{mc}\hat{\mathbf{A}}\displaystyle\sum_{\mathbf{k}\mathbf{q}}e^{i\mathbf{qr}}\hat{c}^{\dagger}_{\mathbf{k}}\hat{c}_{\mathbf{k+q}},
 \label{JFULL}
\end{align}
where  $\hat{c}_{k},\hat{c}^{\dagger}_{k}$ are the  annihilation and creation operators for the electron with momentum $k$, and $\hat{\mathbf{A}}$ is the vector potential operator of the electromagnetic field in second quantized form.  The first term in   Eq.~\eqref{JFULL} is the paramagnetic current which corresponds to  the process of absorption (or emission) of the photon by an electron, which is forbidden for free electrons by the special relativity, since the velocity of the electron after absorption of the photon exceeds the speed of light, thus we will account for the second, diamagnetic term in Eq.~\eqref{JFULL} only. Scattering of the single photon by a single electron is a textbook problem~\cite{Greiner} and the  scattering cross-section is given by:
\begin{align}
\sigma_{\mathrm{TM}}&=r_c^2\iint k_f^2 d k_f d\Omega  \frac{|\mathbf{e}_{k_i}\mathbf{e}_{k_f}|^2}{k_{0i}k_{0f}} \nonumber \\
&\times\delta\left(\frac{\omega_{el}+\Delta\omega}{c}-\sqrt{\left(\frac{\omega_{el}}{c}\right)^2+\Delta k^2}\right), \label{cross_sec_noloss}
\end{align}
where $\omega_{el}=mc^2/\hbar$, $\Delta\omega=\omega_i-\omega_f$, $\Delta k = \sqrt{k_i^2+k_f^2-2k_ik_f\cos\theta}$,$k_i$ and $k_f$ are initial and final electron momenta, $d\Omega=\sin\theta d\theta d \phi$ is the differential of the solid angle, $r_c=e^2/(mc^2)$ is the classical electron radius, and  delta function in the expression Eq.~\eqref{cross_sec_noloss} {ensures} the energy conservation in the system and defines the scattered photon frequency $\omega_f$.
\begin{figure}[!h]
\centerline{\includegraphics[width = 0.8\columnwidth]{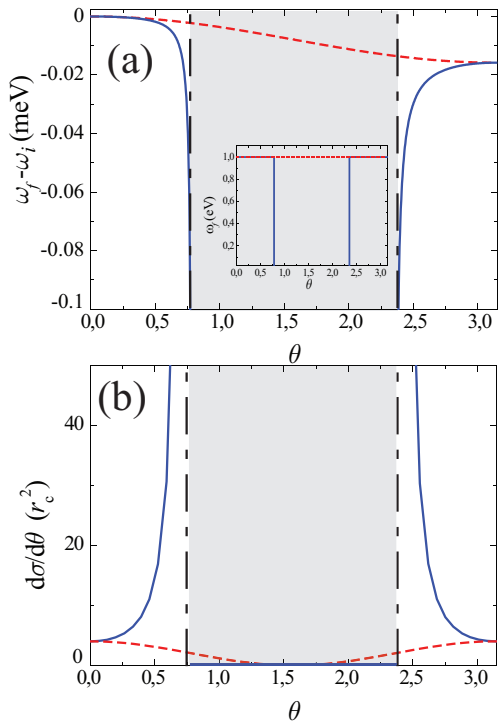}}
\caption{(Color online) Scattered photon frequency (a) and differential cross section (b) for the Compton scattering in lossless dielectric with dielectric permittivity 4 (red dashed lines) and hyperbolic media with $\varepsilon_{xx}=4,\varepsilon_{zz}=-4$ (blue solid lines). }
\label{fig_2}
\end{figure}
The plots of frequency shift and differential cross-section in uniform lossless hyperbolic medium are shown  in Fig.~\ref{fig_2}(a,b). We can see that as the scattering angle approaches $\theta_{cr}=\mathrm{arccot}(\sqrt{\varepsilon_{xx}/|\varepsilon_{zz}}|)$ the scattered photon frequency approaches zero and differential cross-section diverges. The scattering cross-section is exactly zero for the scattering angles $\theta_{cr}<\theta<\pi-\theta_{cr}$. This can be easily explained  if we recall that for these angles the Green's function in lossless hyperbolic medium is  evanescent~\cite{Podd_Green}.

While the consideration of the uniform lossless hyperbolic media gives a good insight into the mechanisms of the free electron scattering in metamaterial,  in order to regularize the divergences occurring in the  expression Eq.~\eqref{cross_sec_noloss} for the differential cross-section and to estimate the possibility of observing the effect experimentally we should consider a more realistic case of the lossy and bounded metamaterial.
{The Fermi Golden rule technique is not applicable due to the presence of losses which should be accounted for via an introduction of an infinite number of additional degrees of freedom. Instead, we use the
Langevin approach when the problem is fully characterized by the Green function determining the electromagnetic response to the local currents~\cite{Welsch}. The calculation procedure involves three steps: (i) calculation of the wave, transmitted inside the medium; (ii) determination of the local currents arising due to the wave scattering on the electrons and (iii) reconstruction of the scattered wave from the known currents. Similar approach has been recently applied to the Brillouin scattering of exciton-polaritons in semiconductor superlattices~\cite{Jusserand}.}

First, we consider the plane electromagnetic wave normally incident upon the interface of semi-infinite hyperbolic medium.  The vector potential of the incident wave in hyperbolic medium can be written as
\begin{align}
\mathbf{A}(z,t)=\mathbf{e_0}{A_0}t_0 e^{-i\omega_{i}t+ik_{0z}^{h}z},\label{eq:A0}
\end{align}
where $\mathbf{e}_0$ is the polarization basis vector,  $\omega_{i}$ is the initial photon frequency,  $t_0$ is the transmission coefficient for the normal incidence, and $k_{0z}^h=\sqrt{\varepsilon_{xx}}\omega_{i}/c$.

Second, in order to consider the scattering we account for the incident field interaction with  an ensemble of free electrons inside the hyperbolic media.
{In experiment the free electrons in hyperbolic media can be generated using the pump-probe technique~\cite{Zayats}.
The incident wave Eq.~\eqref{eq:A0} excites the current  described by a following density operator $\hat{\mathbf j}(\mathbf{r},t)={e^2}/({mc^2})\hat{\rho}(\mathbf{r},t)\mathbf{A}(z,t)$~\cite{Bruus}, that corresponds to the second, diamagnetic, term in Eq.~\eqref{JFULL}.
Here $\hat{\rho}=\frac{1}{V}\sum_{\mathbf{q}_i,\mathbf{q}_f}\hat{c}_{\mathbf{q}_f}^{\dagger}\hat{c}_{\mathbf{q}_i}e^{i(\mathbf{q}_i-\mathbf{q}_f)\mathbf{r}-i(\omega_i-\omega_f)t}$ is the density matrix of the electrons and $V$ is the normalization volume.
{It is the diamagnetic current that produces the scattered wave}.
 The (scattered) electric field operator at position $z_0$ reads
\begin{align}
\hat{\mathbf{E}}(\bm{r},t)=\frac{1}{c^2}\int \bar{\bar{G}}(\bm{r}, \mathbf{r}^{\prime},t-t^{\prime}) \frac{\partial \hat{\mathbf{j}}(\mathbf{r}^{\prime},t^{\prime})}{\partial t^{\prime}}d^3\mathbf{r}^{\prime}dt^{\prime}.
\end{align}
where $\bar{\bar{G}}$ is a dyadic Green's function for the semi-infinite hyperbolic medium. In the case, when the source is inside the hyperbolic medium, and the observation point is outside, at the distance $z_0$ from the interface, the Green tensor in the frequency representation reads~\cite{Novotny}:
\begin{align}
&\bar{\bar{G}}(\bm{\rho},z_0, \bm{\rho}^{\prime},z^{\prime},\omega)=
\int \frac{id^2\mathbf{k}_{\rho}}{(2\pi)^{3}k_{z}}
e^{i\mathbf{k}_{\rho}(\bm{\rho}-\bm{\rho}^{\prime})-i(k_{z}z^{\prime}+k_{z}^{V}z_0)}\nonumber\\&\times\displaystyle
\sum_{\lambda=\rm TE,TM}t_{hV}^{\lambda}\mathbf{e}_{h,\lambda}\otimes\mathbf{e}_{V,\lambda},
\end{align}
where $k_{z}=[\varepsilon_{xx}(\omega/c)^2-(\varepsilon_{xx}/\varepsilon_{zz}){k}_{\rho}^2]^{1/2}$,     $k_{z}^{V}=[(\omega/c)^2-{k}_{\rho}^2]^{1/2}$ is the normal component of the wave vector in vacuum $t_{hV}^{\sigma}$ is the transmission coefficient, and $\mathbf{e}_h, \mathbf{e}_V$ are the polarization basis vectors in hyperbolic media and vacuum respectively.

Finally, performing the  spatial and time Fourier transformation  we obtain the intensity of scattered light,  averaged over the electron distribution
\begin{align}
I(\mathbf{k}_{\rho},z_0,\omega)=\int dt e^{i\omega t} \langle\hat{\mathbf{E}}(\mathbf{k}_{\rho},z_0,t)\cdot \hat{\mathbf{E}}^{\dagger}(\mathbf{k}_{\rho},z_0,0)\rangle.
\end{align}
{The resulting correlators have the form $\mathcal{P}=\langle \hat{c}_{\mathbf{q}_f}^{\dagger}\hat{c}_{\mathbf{q}_i}\hat{c}_{\mathbf{q}_{i^{\prime}}}^{\dagger}\hat{c}_{\mathbf{q}_{f^{\prime}}}\rangle$ and can be readily evaluated using} the Wick's theorem:
\begin{align}
\mathcal{P}=\delta_{ff^{\prime}}\delta_{ii^{\prime}}n_i(1-n_f)+\delta_{if}\delta_{i^{\prime}f^{\prime}}n_in_{i^{\prime}}.
\end{align}
The second term corresponds to the Rayleigh scattering with the conservation of frequency and in-plane wave vector. We focus on the first term, {responsible for the Compton scattering}. The expression for the spectral density function can be written as
\begin{align}
I(\omega,k_{\rho})=E_0^2\left(\frac{\omega}{\omega_{i}}\right)^2\frac{16\pi t_0^2 r_c^2 V^{2/3} m}{\varepsilon_{xx}\hbar}\mathcal{I}(\omega,{k}_{\rho}), \label{dimpart}
\end{align}
where the additional factor of 2 appears from the summation over electron spins, and  the dimensionless factor $\mathcal{I}$ is given by
\begin{align}
&\mathcal{I}(\omega,{k}_{\rho})=\frac{e^{-2\mathrm{Im}k_{z}^Vz_0}}{\left|{1+\frac{k_{z}}{k_{zv}\varepsilon_{\rho}}}\right|^2}
\int\frac{d^3 q_i n_i(1-n_{f(i)})}{|\tilde{q}_{fz}|} \times \nonumber \\
&\displaystyle\sum_{\xi=\pm 1}\frac{1}{[q_{iz}+(k_{0z}-\mathrm{Re}k_{z})+\xi\tilde{q}_{fz}]^2+\mathrm{Im}k_{z}^2}, \label{SpectrDens}
\end{align}
and
\begin{align}
\tilde{q}_{fz}=\sqrt{q_{iz}^2+\frac{2m}{\hbar}(\omega-\omega_i)-k_{\rho}^2+2q_{i\rho}k_{\rho}\cos(\delta \phi)},
\end{align}
where $\delta \phi$ is the angle between the in-plane wave vector of incident radiation and in-plane {wave vector} of electron, $n_{i,f}$ are the Fermi distribution functions and $n_{f(i)}$ means that the electron final energy is determined by the conservation laws:
\begin{align}
E^{el}_f=\frac{\hbar^2}{2m}\left[\tilde{q}^2_{fz}+(\mathbf{q}_{i\rho}-\mathbf{k}_{\rho})^2\right].
\end{align}
The dimension of $I$ in Eq.~\eqref{SpectrDimenless} is volume energy density per frequency per two-dimensional wave vector, thus it corresponds to the intensity of the reflected field at the fixed frequency and wave vector.

Integral in Eq.~\eqref{SpectrDens} can be taken analytically in a special case of zero temperature and small carrier concentration, when the Fermi wave vector $k_F$ is much less than  $\Delta=\sqrt{2m(\omega_i-\omega)/\hbar}$.
{ For the Compton shifts $\hbar(\omega_{i}-\omega)$ as small as  $0.01$ meV this regime is realized up to the  electron densities $n_{el}\approx 4\times 10^{15}~$cm$^{-1}$ and  {$k_F\approx 1.5\times10^5$~cm$^{-1}$.
Below we demonstrate that the actual values of Compton shift can be much larger, so the low density approximation works even better. The dimensionless integral in Eq.~\eqref{SpectrDens} is then given by:}
\begin{align}
\frac{1}{|\sqrt{\Delta^2-{k}_{\rho}^2}|}\displaystyle\sum_{\xi=\pm 1}\frac{4\pi k_F^3/3}{({k}_{0z}-{k}_{zz}+\xi\sqrt{\Delta^2-{k}_{\rho}^2})^2+\mathrm{Im}{k}_{zz}^2}. \label{SpectrDimenless}
\end{align}

\begin{figure}[!h]
\centerline{\includegraphics[width = 0.9\columnwidth]{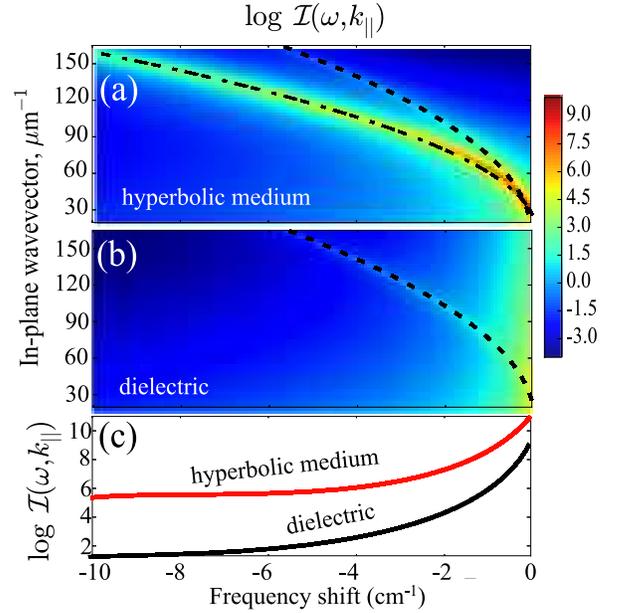}}
\caption{(Color online) (a,b) Logarithmic map of the dimensionless part of the spectral density $I$ given by Eq.~\eqref{SpectrDens} vs in-plane wave vector and  frequency shift for the case of hyperbolic medium (a) and isotropic dielectric (b). (c) Integral of the spectral densities over the in-plane momenta.}
\label{fig_3}
\end{figure}

In Fig.~\ref{fig_3} we plot Eq.~\eqref{SpectrDens} for the case of hyperbolic media with $\varepsilon_{xx}=4.0$, $\varepsilon_{zz}=-4.0+0.3i$ [Fig.~\ref{fig_3}(a)] and for the case of dielectric with the permittivity $4.0+0.3i$ [Fig.~\ref{fig_3}(b)]. {Figure~\ref{fig_3}(c) shows the integrated spectral intensity as function of the Compton shift for both dielectric and hyperbolic regimes.} We do not account for the dielectric permittivity dispersion, since we are working in a very narrow frequency range. The intensity of the scattered radiation is calculated at the distance of $20$~nm from the surface of the structure. The Fermi wave vector is equal to {$10^5$~cm$^{-1}$}.

The spectra of the scattered radiation, shown in Fig.~\ref{fig_3} and calculated numerically after Eq.~\eqref{SpectrDens}, are well approximated by analytical Eq.~\eqref{SpectrDimenless}. The singularities in Eq.~\eqref{SpectrDimenless} describe different scattering resonances. Two scattering channels can be distinguished in Figs.~\ref{fig_3}(a,b).
First channel is indicated by the {dashed} branch  and corresponds to the pole of the first factor of Eq.~\eqref{SpectrDimenless}. The Compton shift is given by $\omega_{i}-\omega_{f}=\hbar k_{\rho}^{2}/(2m)$, i.e.  the scattered electron propagates along the interface with vacuum. This interface feature is similar to the Van Hove singularity in the electronic density of states of one-dimensional systems~\cite{Kittel} or to the Wood anomalies in the grating diffraction~\cite{Hessel}. While it extends up to the large wave vectors, it is integrable and does not contribute much to the overall scattered intensity.

For the hyperbolic media, another scattering channel becomes dominant, marked by a dash-dotted branch in Fig.~\ref{fig_3}(b). This  channel corresponds to the resonance in the second factor of Eq.~\eqref{SpectrDimenless}, described by the bulk momentum conservation law. Large in-plane wave vectors of scattered photons and electrons mean larger scattered electron kinetic energy and, as a result, a larger Compton shift. For dielectric media, this scattering channel is closed because the photon modes with large in-plane wave vectors are not available, see Fig.~\ref{fig_1}.  Figure~\ref{fig_3}(c) demonstrates the main result of our study: the scattering becomes much stronger in the hyperbolic regime so that the Compton shifts up to $\sim 10$cm$^{-1}\sim 1~$meV can be attained.

Hyperbolic metamaterials could facilitate detection of very weak nonlinear processes, as rough plasmonic surfaces already enabled detection of Raman Scattering via manipulation of the local field enhancement. Giant Compton shifts and enhanced scattering cross-sections could be detected via near-field detections, ether inside the bulk, or close to the surface. This statement also relates to any other possible nonlinear scattering processes. We would also like to discuss the feasibility of the effective medium approximation. Internal structure of the hyperbolic media leads to the local field corrections, i.e. that the value of the Compton shift and cross-section depend on the exact position of the electron. However, if the spatial distribution of the free electrons is uniform (which is reasonable approximation if the free electrons are excited with an external  pump), than the local field corrections can be averaged and the effective medium approximation is applicable~\cite{WirePurc}.

In conclusion, we have developed an universal theory for describing nonlinear processes in hyperbolic metamaterials. In particular, we have demonstrated theoretically that the Compton scattering differs drastically in hyperbolic media in comparison with vacuum, and we have predicted that Compton shifts for the visible light frequencies can be enhanced enough to be observed in experiment. The concept of inelastic scattering enhancement in hyperbolic media is rather general, and it can be extended to other quasiparticles, including polaritons, phonons, and surface plasmons.

The work has been supported by the Ministry of Education and Science of the Russian Federation, the Russian Foundation
for Basic Research,  and the Australian Research Council through the ARC Center of Excellence CUDOS.

\end{document}